\documentclass[preprint,12pt]{elsarticle}
\usepackage{amssymb}
\usepackage{graphicx,amsmath}
\usepackage{graphicx}
\usepackage{amsmath}
\usepackage{color}
\usepackage{enumerate}
\usepackage{changes}

\journal{Astroparticle Physics}

\begin{document}
\begin{frontmatter}
\title{Searching for Gravitational Waves with a \\Geostationary Interferometer}

\author[inpe]{Massimo Tinto\fnref{fn1}}
\ead{Massimo.Tinto@jpl.nasa.gov}
\author[inpe]{Jose C.N. de Araujo}
\ead{jcarlos.dearaujo@das.inpe.br}
\author[inpe]{Odylio D. Aguiar}
\ead{odylio.aguiar@inpe.br}
\author[unifei]{M\'arcio E.S. Alves\corref{cor1}}
\ead{alvesmes@unifei.edu.br}
\address[inpe]{Divis\~{a}o de Astrof\'{i}sica, Instituto Nacional de Pesquisas Espaciais, S. J. Campos, SP 12227-010, Brazil}
\address[unifei]{Instituto de F\'isica e Qu\'imica, Universidade Federal de Itajub\'a, Itajub\'a, MG 37500-903, Brazil}

\cortext[cor1]{Corresponding author}
\fntext[fn1]{Also at: Jet Propulsion Laboratory, California Institute of Technology, Pasadena, CA 91109}

\begin{abstract}
  We analyze the sensitivities of a geostationary gravitational wave
  interferometer mission operating in the sub-Hertz band. Because of
  its smaller armlength, in the lower part of its accessible frequency
  band ($10^{-4} - 2 \times 10^{-2}$ Hz) our proposed Earth-orbiting
  detector will be less sensitive, by a factor of about seventy, than
  the Laser Interferometer Space Antenna (LISA) mission.  In the
  higher part of its band instead ($2 \times 10^{-2} - 10$ Hz), our
  proposed interferometer will have the capability of observing
  super-massive black holes (SMBHs) with masses smaller than $\sim
  10^{6}$ M$_{\odot}$. With good event rates for these systems, a
  geostationary interferometer will be able to accurately probe the
  astrophysical scenarios that account for their formation.
\end{abstract}


\end{frontmatter}

\section{Introduction}
\label{intro}

The quest for the direct observation of gravitational radiation is one
of the most pressing challenges in the physics of this century.
Predicted by Einstein shortly after formulating his general theory of
relativity, gravitational waves (GW) will allow us to probe regions of
space-time otherwise unobservable in the electromagnetic spectrum
\citep{Tho87}.  Although we have excellent indirect evidence of their
existence through their effects on the orbital evolution of binary
systems containing pulsars, we have not been able yet to directly
detect them.  Several ground-based gravitational wave detectors have
been operational for awhile, and only recently they have been able to
identify the most stringent upper limits to date for the amplitudes of
the radiation expected from several classes of sources
\citep{LIGO,VIRGO,GEO,TAMA}. Next generation of Earth-based
interferometers and pulsar-timing experiments \citep{PULSAR}, as well
as newly conceived space-born detectors \citep{PPA98} are expected to
achieve this goal.

Ground-based interferometers operate in the frequency band whose lower
limit is at about $10$ Hz, mainly because of the large seismic and
gravity-gradient noises below this frequency cut-off.  Since the mHz
region is potentially very rich in gravitational wave sources, the
most natural way to observe them is to build and operate a space-borne
detector. The most notable example of a space interferometer, which
has been under study for several years jointly by scientists in the
United States of America and in Europe, is the Laser Interferometer
Space Antenna (LISA) mission. By relying on coherent laser beams
exchanged between three remote spacecraft forming a giant (almost)
equilateral triangle, LISA aimed to detect and study cosmic
gravitational waves in the $10^{-4} - 1$ Hz band.

Although over the years only a few space-based detector mission
concepts have been considered as alternatives to the LISA mission
(with ASTROD \citep{ASTROD}, DECIGO \citep{DECIGO}, and OMEGA
\citep{Hiscock} being the most notable examples), starting in 2011
(with the ending of the NASA/ESA partnership for flying LISA) more
mission concepts have appeared in the literature \citep{PCOS}. Their
goals are to meet most (if not all) the LISA scientific objectives
(highlighted in the 2010 Astrophysics Decadal Survey {\it New Worlds,
  New Horizons (NWNH)} \citep{NWNH}) at a lower cost.

In this context, motivated by the interest of the Brazilian Space
Agency to pursue the development and launch of several geostationary
satellites, we decided to analyze the scientific capabilities offered
by a geostationary gravitational wave interferometer, henceforth
called GEOGRAWI. GEOGRAWI was one of the alternative concepts to the
LISA mission submitted in response to NASA's Request for Information
\# NNH11ZDA019L \citep{RFI}. A detailed analysis of all the submitted
projects and a final study report can be found in \citep{PCOS}. Among
them, a proposal by Sean T. McWilliams explores a mission concept very
similar to the one discussed in this paper \citep{OURRFI,Sean2011}
called GADFLI. His concept and the one proposed by us, though
strikingly similar, were nonetheless independently developed and
submitted.

GEOGRAWI, like LISA, has three identical spacecraft interchanging
coherent laser beams and forming an (almost) equilateral triangle.
However, by being in a geostationary orbit, its
accessible frequency band changes to about ($10^{-3} - 10$ Hz).
It is worth noting that, although our interferometer reminds that of
the OMEGA mission (proposed by R.  Hellings \citep{Hiscock} about 15
years ago), it is in fact very different from it. The OMEGA spacecraft
were six in total, with two at each vertex of a geocentric equilateral
triangle of side equal to one million kilometers. Cost of the
launching vehicle as well as challenges in performing the phase
measurements associated to the large relative velocities between the
spacecraft make OMEGA quite different from GEOGRAWI.

The astrophysical sources that GEOGRAWI is expected to observe within
its operational frequency band include extra-galactic massive and
super-massive black-hole coalescing binaries, the resolved galactic
binaries and extra-galactic coalescing binary systems containing white
dwarfs and neutron stars, a stochastic background of astrophysical and
cosmological origin, and possibly more exotic sources such as cosmic
strings. With GEOGRAWI we will be able to test Einstein's theory of
relativity by comparing the waveforms detected against those predicted
by alternative relativistic theories of gravity, and also by measuring
the number of independent polarizations of the detected gravitational
wave signals \citep{TintoAlves2010}.

Since the sensitivities of the geostationary interferometers we
considered (see section \ref{SENSITIVITY} below) are degraded by their
shorter armlength in the frequency region ($10^{-4} - 2 \times
10^{-2}$) Hz, it will be impossible for them to detect the zero-order
cyclic spectrum of the white dwarf-white dwarf galactic binary
confusion noise. However, as pointed out in \citep{Edlund}, the
periodic nature of the galactic background signal in the data of an
interferometer rotating around the Sun will contain higher-order
``cyclic spectra'', which are in principle not affected by the
instrumental noise ({\underbar {if}} this is stationary). This implies
that we could still detect the so-called ``confusion noise'', and
infer properties of the distribution of the white-dwarf binary systems
present in our galaxy \citep{Edlund}.

In this article we pay particular attention to super-massive black
holes, which are the main sources of GWs for interferometers operating
at frequencies below $1$ Hz. We show that GEOGRAWI will be able to
detect a large number of SMBHs at very high redshifts, and that a
significant fraction of them will have masses smaller than $10^{6}$
M$_{\odot}$ as a consequence of its improved sensitivity at higher
frequencies.

This paper is organized as follows. In section \ref{SENSITIVITY} we
provide a description of the mission and a system noise analysis in
order to evaluate the sensitivities of the Time-Delay Interferometric
(TDI) \citep{TD2005} response $X$. It is in this section and in
\ref{noise} where we also emphasize that additional instrumental
techniques over those baselined for LISA will need to be implemented
in order to achieve the derived sensitivities. These are estimated for
three different on board subsystem configurations of GEOGRAWI, which
we refer to as Geostationary 1 (GEO1), Geostationary 2 (GEO2) , and
Geostationary 3 (GEO3). We note that the GEO1 configuration has a
reduced sensitivity over that of LISA by a factor of about $70$ in the
lower part of its accessible frequency band ($10^{-4} - 2 \times
10^{-2}$ Hz). On the other hand, in section \ref{Science} we discuss
possible sources of gravitational waves observable by GEOGRAWI in the
higher-part of its band ($2 \times 10^{-2} - 10$ Hz). Particular
attention is devoted to SMBHs and we show that the number of
observable signals from such systems, with masses smaller than or
equal to $10^6$ M$_\odot$, would be significant. Finally, in Section
\ref{Conclusions} we provide a summary of our work and conclusions.

\bibliographystyle{model1a-num-names}

\section{Mission Design and Interferometer Sensitivities}
\label{SENSITIVITY}

Our proposed space-based detector entails three spacecraft in
geostationary orbit, forming an equilateral triangle with armlength of
about $73,000$ km.  Since the constellation plane coincides with the
Earth equatorial plane, each spacecraft will need to be spherical and
entirely covered with photo-voltaic cells. This will allow each
spacecraft to be continuously powered (without the need of rotating it
as the Earth orbits around the Sun) and maintain a high-level of
thermal stability. It should be noticed also that the inclination
($23.5^0$) of the Earth equatorial plane with respect to the ecliptic
will prevent (i) Sun light from contaminating the optics of the
spacecraft during most of the interferometer orbital period around the
Sun, and (ii) spacecraft occultation by the Earth. In proximity of the
equinoxes, however, these properties will no longer be true and, in
order to avoid these short-period operation interruptions, onboard electric
power and Sun-light filters will need to be used.  Sun-light filters
needed for this purpose have already been developed for the OMEGA
mission, which requires their usage at all times in its geocentric
trajectory \citep{RONRFI}.

As the Sun and the Moon will exercise gravitational perturbations on
each spacecraft (resulting into a long-term orbital drift), in order
to maintain orbital stability and small inter-spacecraft relative
velocities, each spacecraft will need to implement ``station-keeping''
\citep{Soop}. This is required not only for keeping the constellation
in a stable configuration, but most importantly for relying on a
phase-meter design that does not need to accommodate large relative
frequency offsets, making it significantly less noisy (see
\ref{noise}). By taking the nominal value of $2 \ {\rm m/s/year}$ for
the ``East-West'' acceleration perturbation \citep{Soop} acting on the
spacecraft (which is responsible for the spacecraft relative
velocities), it is easy to show that, in order to maintain their
maximum velocities smaller than about a few decimeters per second, the
station-keeping operation will need to be applied about once per week
(see \ref{noise}).

In order to estimate the TDI sensitivities of our geostationary GW
interferometer detector, we have relied on some LISA study
documents \citep{PPA98,LISA1}. This allowed us to derive a break-down estimate
of the various noise sources affecting the TDI observables and compute
the TDI sensitivities of a geostationary interferometer under the
following three different on-board subsystem configurations:

\begin{enumerate}[(I)]
\item Some of the onboard science payload components (the laser, the
  optical telescope, the inertial reference sensor) are assumed to be
  equal and have a similar noise performance of those planned to be
  flown onboard the LISA mission. Other subsystems are regarded to
  have a noise performance that results in a high-frequency noise
  spectrum that is essentially determined by the photon-counting
  statistics. It should be emphasized, however, that our assumed noise
  performance in the high-frequency region of the accessible band
  requires additional instrumental developments over those baselined
  for the LISA mission \cite{Jordan2006,Dan1,Dan2}. In order to better
  understand this point we refer the reader to \ref{noise} where
  it is discussed somewhat at length. We will refer to this
  configuration as ``Geostationary 1'' (or GEO1).

\item The output power of the onboard lasers and the size of the
  optical telescopes are assumed to be equal to those of the LISA
  mission, while the noise performance of the gravitational reference
  sensors (GRS) is taken to be a factor of ten worse than that of the
  GRSs planned for the LISA mission. The remaining noises
  are as in the GEO1 case, and this configuration will be referred to
  as ``Geostationary 2'' (or GEO2).  \footnote{This GRS
    noise level is equal to that of the GRS to be flown
    on-board the LISA Pathfinder mission \citep{Vitale1,Vitale2}}

\item The noise performance of each GRS is taken a factor of
  ten worse than that of the GRS planned for the LISA
  mission, the output power of the lasers is assumed to be a factor of
  $10$ smaller than that of the lasers onboard LISA, the diameter of
  the optical telescopes has been reduced by a factor of $\sqrt{10}$
  over that of the LISA telescopes, and the remaining noises are
  assumed to be equal to those for the GEO1 configuration. This
  configuration will be called ``Geostationary 3'' (or GEO3).
\end{enumerate}

\begin{figure}
\begin{center}
\includegraphics[width = 6 in]{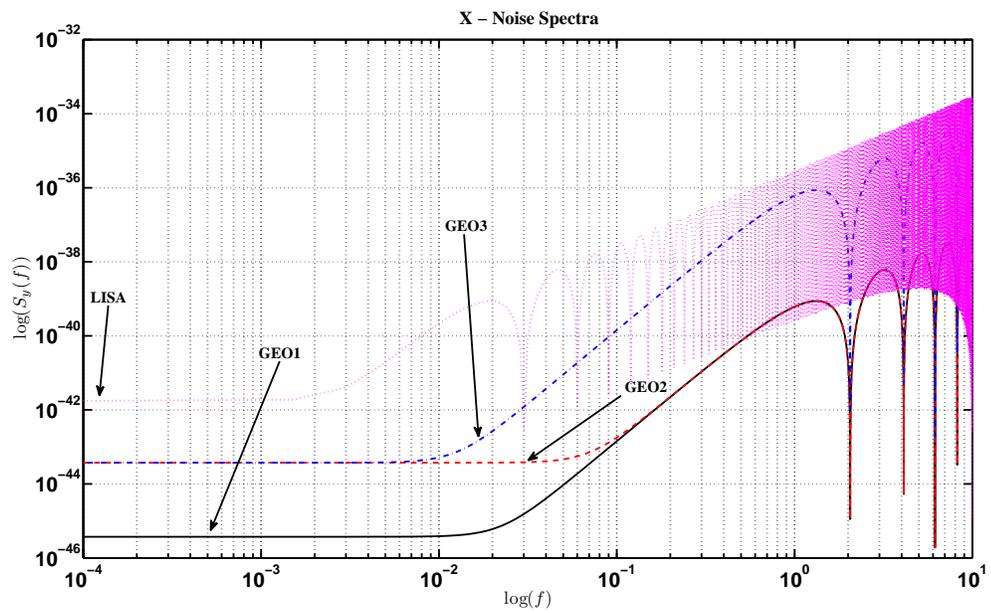}
\caption{Noise spectra for the $X$
  time-delay interferometric combination for the three on-board
  hardware configurations (I), (II), and (III) (see text).  The
  varying depths of the minima in the high-frequency range is an
  artifact of numerically calculating these functions at discrete
  frequencies}
\label{NoiseGEO}
\end{center}
\end{figure}

The sensitivity of an interferometer detector of gravitational
radiation has been traditionally taken to be equal to (on average over
the sky and polarization states) the strength of a sinusoidal GW
required to achieve a signal-to-noise ratio of $5$ in a one-year
integration time, as a function of Fourier frequency.  For sake of
simplicity we will limit the derivation of the sensitivities of the
three geostationary configurations described above to only the ``X''
TDI combination, as those for the other TDI combinations can be
inferred by properly scaling those for $X$. To this end we will be
following the same procedure described in \citep{AET1999,ETA2000}, and
rely on the following expression for the power spectrum of the noises
affecting the $X$ combination (see Figure \ref{NoiseGEO})
\begin{equation}
S_X (f) = [8 \sin^2(4 \pi f L) + 32 \sin^2(2 \pi f L)] S^{pm}_y (f) +
16 \sin^2(2 \pi f L)\ S^{op}_y (f) \ ,
\label{eq:10}
\end{equation}
where $S^{pm}_y (f)$ is the spectrum of the relative frequency
fluctuations due to each proof mass, and $S^{op}_y (f)$ is the
spectrum of optical path (mainly shot and beam pointing) noise.  Both
these noises have been regarded as the main limiting noise sources for
LISA \citep{PPA98,ETA2000}, and will be treated as such for our
geostationary interferometer configuration (see \ref{noise} for
details). It should be acknowledged, however, that the dramatic
reduction in optical path noise assumed in this article reflects the
implementation of additional instrumental techniques over those
baselined for LISA \cite{Jordan2006,Dan1,Dan2}.

The numerical expressions for the spectra $S^{pm}_y (f)$, $S^{op}_y
(f)$ associated to the LISA mission were provided in \citep{ETA2000}
and they are equal to $S^{pm}_y (f)|_{LISA} = 2.5 \times 10^{-48} \ [f
/1 Hz]^{-2} \ {\rm Hz}^{-1}$ and $S^{op}_y (f)|_{LISA} = 1.8 \times
10^{-37}[f /1 Hz]^2 \ {\rm Hz}^{-1}$. The configurations (I), (II) and
(III) discussed above require appropriate scaling factors of these
noise levels, and they are given in Table \ref{Table1} (also see
\ref{noise} for detail analysis of the other noise sources).

\begin{table}[h!]
\begin{center}
\begin{tabular}{| c | c | c |}
\hline
Mission &  $S^{op}_{y}|_{LISA}$ & $S^{pm}_{y}|_{LISA}$  \\
\hline
LISA            & $1$ & $1$ \\
\hline
Geo. 1         & $(L_{GEO}/L_{LISA})^2$ & $1$  \\
\hline
Geo. 2 & $(L_{GEO}/L_{LISA})^2$ & $10^2$  \\
\hline
Geo. 3 & $(L_{GEO}/L_{LISA})^2 \times 10^3$ & $10^2$  \\
\hline
\end{tabular}
\end{center}
\caption{Noise spectra scaling factors as functions of the different
  instrument configurations analyzed. See text for more details.}
\label{Table1}
\end{table}

Gravitational wave sensitivity is the wave amplitude required to
achieve a given signal-to-noise ratio.  We calculate it in the
conventional way, requiring a signal-to-noise ratio of $5$ in a one
year integration time: $5 \ \sqrt{S_X(f) \ B}$/(root-mean-squared
gravitational wave response of $X$). The bandwidth, $B$, was taken to
be equal to one cycle/year (i.e. $3.17 \times 10^{-8}$ Hz), while  the
root-mean-squared gravitational wave response was calculated by
averaging over random sources of monochromatic GWs uniformly
distributed over the celestial sphere and over their polarization
states. This was done by taking  the wave functions, ($h^{(+)}$, $h^{(\times)}$),
in terms of a nominal wave amplitude, $H$, and the two
Poincar\'e parameters, ($\Phi, \Gamma)$, in the following way \citep{AET1999}
\begin{eqnarray}
h^{(+)} (t) & = & H \ \sin(\Gamma) \ \sin(\omega t + \Phi) \ ,
\label{h2}
\\
h^{(\times)} (t) & = & H \ \cos(\Gamma) \ \sin(\omega t) \ .
\label{h3}
\end{eqnarray}
We averaged over source direction by assuming uniform distribution of
the sources over the sky, and also averaged over elliptical
polarization states uniformly distributed on the Poincar\'e sphere for
each source direction.  The averaging was done via Monte Carlo
integration with $10000$ source position/polarization state pairs per
Fourier frequency bin and $10000$ Fourier bins across the entire band
analyzed ($10^{-4} - 10$) Hz.

\begin{figure}
\begin{center}
\includegraphics[width = 6 in]{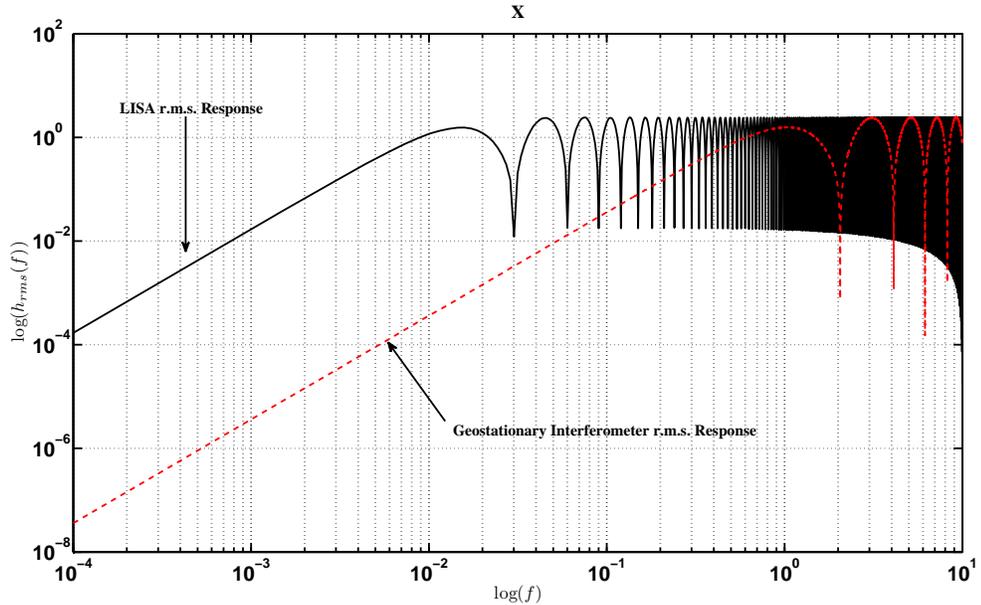}
\caption{Root-Mean-Square response of the $X$ TDI combination for a
  geostationary interferometer. For completeness we have included also
  that for the interplanetary LISA mission, whose armlength is of $5 \
  \times 10^6$ km. The r.m.s. has been calculated by assuming an
  ensemble of sinusoidal signals uniformly distributed on the
  celestial sphere and randomly polarized.}
\label{RMS}
\end{center}
\end{figure}

Figure \ref{RMS} shows the root-mean-squared (r.m.s.) responses of
the TDI combination $X$ for a geostationary interferometer and for the
interplanetary LISA mission (which is shown here for comparison). In
the low-part of the accessible frequency band we may notice that the
r.m.s.  response of the geostationary interferometer is, as expected,
penalized by the shorter armlength. At higher frequencies instead it is
comparable to that of LISA because, in this region of the band, the
response does no longer grow with the armlength of the detector. Note
also that the varying depths of the minima in the high-frequency range
is an artifact of numerically calculating these functions at discrete
frequencies.

\begin{figure}
\begin{center}
\includegraphics[width = 6 in]{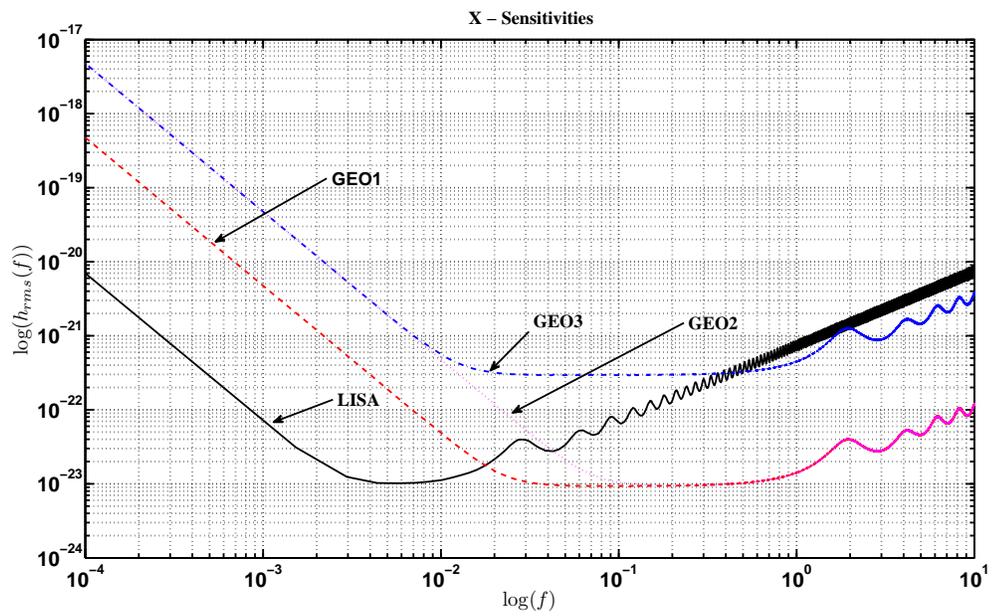}
\caption{Sensitivity of the $X$ combination of a geostationary
  interferometer. Its noise performance is characterized by the noise
  spectra shown in Table (\ref{Table1}) associated to the three
  different onboard subsystems configurations discussed. See text for
  more details.}
\label{Sensitivities}
\end{center}
\end{figure}

In Figure \ref{Sensitivities} we then plot the sensitivities of the
TDI combination $X$ for the various on-board hardware configurations
discussed above. The characteristic behavior of the r.m.s.  responses
in Figure \ref{RMS} folds into the plots presented here. At
high-frequencies ($f \ge 0.05$ Hz) the sensitivities of the GEO1 and
GEO2 configurations are significantly better than that of the LISA
mission, while that of the GEO3 configuration shows a sensitivity
marginally better than that of LISA only for frequencies $f \ge 0.5$
Hz. At lower frequencies instead, the LISA's longer armlength results
into a sensitivity better than those of the GEOGRAWI configurations we
considered.

Although the sensitivities at lower frequencies of any of the
geostationary interferometers we considered are penalized by their
shorter armlength or worse GRS noise, they can still provide
a wealth of information about several astrophysical sources LISA was
expected to detect and study. In the following section we turn to the
science that a GEOGRAWI detector will be able to deliver.

\section{Science with GEOGRAWI}
\label{Science}

In the previous section we have seen that a GEOGRAWI can operate in
the $10^{-4} - 10$ Hz frequency band, and it could be more sensitive
than LISA (configuration (I,II) discussed in section
\ref{SENSITIVITY}) by a factor as large as $70$ in the frequency
region above $20$ mHz. As a consequence of this result, it is natural
to presume that a geostationary detector will be able to observe
massive and super-massive Black Holes (SMBHs) and stellar-mass binary
systems in the region of the frequency band where it achieves its best
sensitivity. In the next subsections we focus our attention on SMBHs,
as the sensitivity plots presented in Section \ref{SENSITIVITY}
already imply the detectability of several binary systems present in
our own galaxy (the so called ``calibrators'') \citep{PPA98}, and other
sources emitting in the higher region of the accessible frequency
band.

\subsection{\label{smbh}Detecting Supermassive Black Holes with GEOGRAWI}

A significant amount of GW energy can be released during the three
evolutionary phases (namely inspiral, merger and ring-down) of
coalescing binaries containing SMBHs (BSMBH). To assess how well and
how often a geostationary interferometer could detect SMBHs during
these three phases, one should calculate the maximum redshift, for a
given signal-to-noise ratio (SNR), at which these systems could be
detectable.

To perform the calculations above one needs to know the waveforms. For
the inspiral and ringdown phases there are well known analytical
solutions \citep[see, e.g.,][]{flhu}, which provide reliable results.
For the merging phase, however, one needs to rely on fully numerical
solutions of Einstein's equations
\citep{PRE2005,BA2006,CA2006,MC2011,HI2010}.\footnote{We refer the
  reader to the review paper by Centrella {\it et al.}  \citep{CE2010}
  where a historical overview of the main highlights in numerical
  relativity, and of the binary black-hole (BBH) simulations in
  particular, are presented together with a detailed bibliography.}

In recent years, it has been shown in the literature
\citep[see, e.g.,][]{Ajith2008} that there exist some reliable
analytic approximations of the numerically derived waveforms valid
under the assumption of two non-spinning black-holes.

In order to calculate the maximum redshift we consider the formula for
the SNR based on the well known matched filtering technique (see, for
example, Flanagan and Hughes \citep{flhu}). The matched-filtering
expression for the averaged squared SNR, in terms of the energy
spectrum of the gravitational waves $dE/df$, is equal to \citep{flhu}

\begin{equation}
\langle SNR^2 \rangle = \frac{2 (1 + z)^2}{5 \pi^2 D_{L}(z)^2} \,
\int_0^\infty df \, \frac{1}{f^2 S_h(f)} \, \frac{d E}{d f}[(1 +
z)f] \ ,
\label{snr}
\end{equation}
where $z$ is the cosmological redshift of the source, $D_{L}$ is the
corresponding luminosity distance and $S_h(f)$ is the spectral density
of the strain noise of the GW detector. On the left-hand-side of Eq.
(\ref{snr}) the angle-brackets denote an ensemble average over sources
uniformly distributed over the celestial sphere and polarization
states.

Once the energy spectrum of the GWs (or its corresponding waveform) is
given, Eq. (\ref{snr}) allows us to infer, for a given signal-to-noise
ratio and a specific GEOGRAWI configuration, the maximum redshift at
which a source can be detected, and then to estimate the SMBHs
observable event rate.

It is worth noting that, although introduced many years before the first
fully numerical relativity calculations of BBH mergers were available,
the waveform proposed by Flanagan \& Hughes \citep{flhu} gives results
that are within an order of magnitude of those derived numerically.
Numerical relativity in fact showed that their white spectrum
approximation for characterizing the merging phase resulted into an
overestimated emitted GW energy \citep[see, e.g.,][]{BM2007}.

In the next subsections we separately consider the three different
phases of the coalescing BSMBHs mentioned above. For the inspiraling
and the ringdown phases we adopt the energy spectrum presented by
Flanagan \& Hughes \citep{flhu}. For the merging phase we adopt
instead the waveform by Ajith {\it et al.} \citep{Ajith2008}, which
captures in analytic form the features of the waveforms derived
numerically for two non-spinning black holes. Some of our
calculations are also based on the analysis recently done by Filloux
{\it et al.}  \citep{fill,fill0} on the formation and evolution of
SMBHs.

\subsubsection{\label{ringdown}Detecting SMBH in the Ringdown phase}

In the present study the SMBHs are formed as a result of the merging
of two other SMBHs, which for simplicity have been assumed of equal
masses. The SMBH so formed goes through the process of ``ringdown'',
and the resulting characteristic wave it emits is a damped sinusoid
\begin{equation}
\label{signal}
h(t)\propto e^{-t/\tau}\cos(2\pi f_{rd}t) \ ,
\end{equation}
where $\tau$ is the characteristic ringdown damping time and $f_{rd}$
is the frequency of the radiated signal. Following, e.g., Ref. \citep{flhu},
the energy spectrum is equal to
\begin{equation}
\frac{dE}{df} \simeq \frac{1}{8} {\cal A}^2 Q M^2 f_{rd} \, \, \delta(f -f_{rd}),
\label{dEdfRD}
\end{equation}
where $M$ is the SMBH mass, ${\cal A}$ is a dimensionless coefficient
that describes the magnitude of the perturbation when the ringdown
begins and $Q$ is the quality factor of the mode.

Following Ref. \citep{tche}, $f_{rd}$ and $Q$ can be related to the
spin parameter $a$ and to the mass $M$ in the following way
\begin{equation}
{f_{rd}} \simeq \left[1-0.63(1-a)^{3/10}\right] \ \frac{1}{2 \pi M},
\label{frd}
\end{equation}
\begin{equation}
Q = \pi\tau f_{rd} \simeq 2 \, (1-a)^{-9/20} \ .
\end{equation}

The energy associated with the ringdown process can be written as

\begin{equation}
E_{\rm rd} \simeq \frac{1}{8} {\cal A}^2 M^2 f_{rd} Q,
\label{rdenergy}
\end{equation}
which we will take to be a fraction, $\varepsilon_{rd}$, of the total
mass $M$ of the black-hole.  Although it has been argued in the
literature that $\varepsilon_{rd}$ could very well be as large as a
few percents (see Ref. \citep{flhu} for an interesting discussion on
this issue), here we will consider only two different (conservative)
values for it, namely: $\varepsilon_{rd} = 1 \ \%$ and
$\varepsilon_{rd} = 0.1 \ \%$.

Finally, from the above equations, the SNR (Eq. \ref{snr}) can be
rewritten in the following form
\begin{equation}
\langle SNR^2\rangle = \frac{8}{5} \, \frac{\varepsilon_{rd}}{D_{L}(z)^2 F(a)^2}
\frac{(1+z)^3 M^3}{S_h[f_{rd}/(1+z)]},
\label{snrrd}
\end{equation}
where $F(a) = 1 - 0.63 (1 - a)^{3/10}$.

With the expression (\ref{snrrd}) in hand we can calculate the redshift,
$z_{max}$, below which a given SMBH of mass $M$ can be detected with a specified
SNR, spin parameter $a$ and energy fraction $\varepsilon_{rd}$.

Since the luminosity distance depends on the cosmological parameters,
in what follows we will assume a $\Lambda$CDM flat cosmology with the
Hubble parameter $H_{0} = 70\, {\rm km}\, {\rm s}^{-1}\, {\rm
  Mpc}^{-1}$, the total matter density parameter $\Omega_{m}=0.3$ and
the cosmological constant density parameter $\Omega_{\Lambda}=0.70$.

\begin{figure}
\begin{center}
\includegraphics[width=10cm]{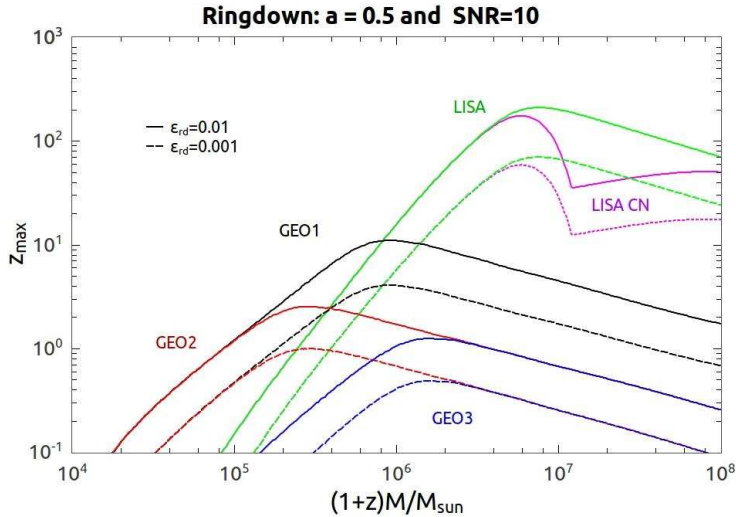}
\caption{$z_{max}$ as a function of the SMBH mass M for $SNR=10$, $a=0.5$ and
  $\varepsilon_{rd}$ = 0.01 and 0.001, for the three GEOGRAWI
  configurations.  Also shown for comparison are the results for LISA
  with (LISA CN) and without (LISA) the confusion noise contribution.}
\label{ring}
\end{center}
\end{figure}

In Figure \ref{ring} we plot $z_{max}$ as a function of the mass of
the SMBH by taking ${\rm SNR}= 10$ for the three GEOGRAWI
configurations discussed in Section \ref{SENSITIVITY}. Also shown for
comparison are the results for LISA with (LISA CN) and without (LISA)
the confusion noise (CN) contribution. Since $z_{max}$ mildly depends
on the spin parameter $a$, we have fixed its value to $0.5$. On the
other hand, since $z_{max}$ is very sensitive to the percentage of radiated
energy, we have analyzed the following two possible values for
$\varepsilon_{rd}$: $0.01$ and $0.001$.

Although LISA can observe these signals at much larger redshifts than
any GEOGRAWI configuration discussed in this article, the GEO1
configuration certainly represents an interesting alternative to LISA
by being able to observe SMBHs at redshifts as high as $z = 10$ when
$\varepsilon_{rd}=0.01$.

From the above results we can now estimate the event rates of these
systems for the different GEOGRAWI configurations. To perform this
calculation we model the formation of SMBHs by relying on a recent
study by Filloux {\it et al.} (see \citep{fill} and references
therein), where the coalescence history of SMBHs is derived from
cosmological simulations. Filloux {\it et al.} were able to derive the
coalescence rate per unit volume and mass intervals as a
function of the resulting BH mass and redshift (see Fig. 3 of
\citep{fill}), $\Psi(M,z)$.  With $\Psi(M,z)$ we can write the
differential coalescence rate at the detector frame in the following
way
\begin{equation}\label{erobs}
dR_{obs}(M,z) = \frac{\Psi(M,z)}{1+z}  dM \frac{dV}{dz}dz \ ,
\end{equation}
where the factor $(1+z)$ takes into account the time dilation.

For additional discussions, concerning the coalescence rate related to
this study and the corresponding cosmological simulations, besides
Ref. \citep{fill} we refer the reader to Ref. \citep{fill0} and \ref{appendix}.

Note that the above equation depends on the co-moving volume element
$dV$ which, for a flat cosmology, is equal to
\begin{equation}
dV=4\pi\left(\frac{c}{H_0}\right)\frac{r^2(z)dz}{\sqrt{\Omega_{\Lambda}+\Omega_m(1+z)^3}}
\end{equation}
\par\noindent and $r(z)$ is the co-moving distance
\begin{equation}
r(z)=\frac{c}{H_0}\int_0^z\frac{dz^{\prime}}{\sqrt{\Omega_{\Lambda}+\Omega_m(1+z^{\prime})^3}}.
\end{equation}

With $z_{max}(M)$ we can now integrate Eq. (\ref{erobs}) in
the $M$ and $z$ variables to estimate the event rate for each GEOGRAWI
configuration as a function of the SNR, $a$ and $\varepsilon_{rd}$.

\begin{figure}
\begin{center}
\includegraphics[width=10cm]{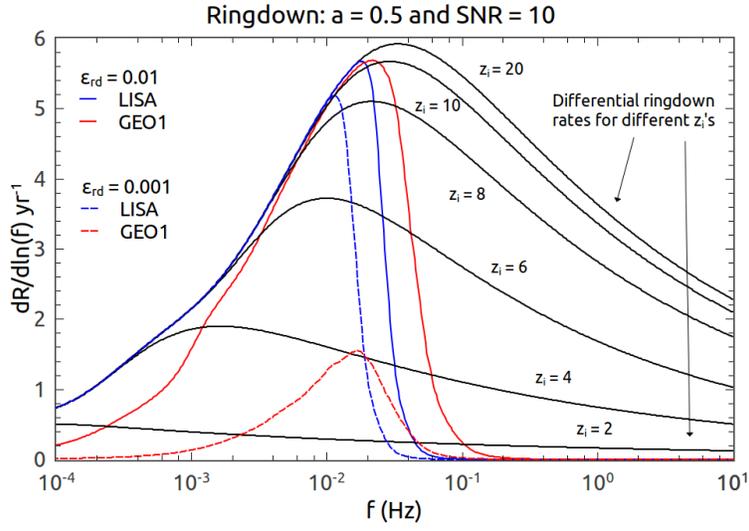}
\caption{
  Differential ringdown observation rate as a function of frequency
  for the GEOGRAWI configuration 1 (GEO1) and for LISA. Also plotted,
  the differential ringdown production rate for different initial
  redshifts (notice that not all of it is seen neither by LISA nor
  GEO1); see \ref{appendix} for details.}
\label{Differ}
\end{center}
\end{figure}

In Table II we show our estimated ringdown observation rates for
GEOGRAWI and, for comparison, also those for LISA. It can be noticed
that, for $\varepsilon_{rd} = 0.01$, the event rates for GEO1 are
comparable to those for LISA. This can be better understood with the
aid of Fig.\ref{Differ}, which shows the logarithmic differential
ringdown observation rate as a function of frequency, together
with the differential ringdown rate for comparison (see
\ref{appendix} for details.)

Note that when $\varepsilon_{rd} = 0.01$ the curve of the differential
ringdown observation rate of GEO1 extends at higher frequencies more
than the LISA curve. In this region of its accessible frequency band
GEO1 can detect ringdown gravitational waves emitted by systems of
smaller masses. Since these observed systems are larger in number than
the more massive SMBHs, we can see why GEO1 has a ringdown observation
rate comparable to that of LISA. On the other hand, as it will be
  shown in section \ref{spiral}, these same systems will be observed
  by LISA with a SNR larger than that achievable by GEO1 during the
  inspiraling phase.

As shown in \ref{appendix}, the model by Filloux et al. \citep{fill}
for the current value of the coalescence rate amounts to $\simeq$ 43
yr$^{-1}$.  This implies that, depending on the values of
$\varepsilon_{rd}$ and $a$, GEO1 can detect a large fraction of
ringdown SMBHs.

\begin{table}
\label{tab2}
\begin{center}
\caption{\label{event}The ringdown observation rates for LISA,
GEO1, GEO2 and GEO3 with a SNR = 10 and for different values
of $\varepsilon_{rd}$ and $a$.}
\begin{tabular}{c c c c c c c}
\hline
Antenna  &  & $a$     &  & $\varepsilon_{rd}$  &  & R(yr$^{-1}$) \\ \hline
LISA CN  & &  0      & & 0.001               & & 16.8         \\
LISA CN  & &  0.5    & & 0.001               & & 15.7         \\
LISA CN  & &  0.95   & & 0.001               & & 14.1         \\
LISA CN  & &  0      & & 0.01                & & 19.8         \\
LISA CN  & &  0.5    & & 0.01                & & 18.5         \\
LISA CN  & &  0.95   & & 0.01                & & 16.7         \\ \hline
GEO1 & &  0      & & 0.001   & & 2.41         \\
GEO1 & &  0.5    & & 0.001   & & 3.27         \\
GEO1 & &  0.95   & & 0.001   & & 5.29         \\
GEO1 & &  0      & & 0.01    & & 16.8         \\
GEO1 & &  0.5    & & 0.01    & & 18.3         \\
GEO1 & &  0.95   & & 0.01    & & 19.0         \\ \hline
GEO2   & &  0      & & 0.001   & & 0.049        \\
GEO2   & &  0.5    & & 0.001   & & 0.066        \\
GEO2   & &  0.95   & & 0.001   & & 0.10         \\
GEO2   & &  0      & & 0.01    & & 0.58         \\
GEO2   & &  0.5    & & 0.01    & & 0.82         \\
GEO2   & &  0.95   & & 0.01    & & 1.35         \\ \hline
GEO3   & &  0      & & 0.001   & & 0.008        \\
GEO3   & &  0.5    & & 0.001   & & 0.011        \\
GEO3   & &  0.95   & & 0.001   & & 0.018        \\
GEO3   & &  0      & & 0.01    & & 0.098        \\
GEO3   & &  0.5    & & 0.01    & & 0.13         \\
GEO3   & &  0.95   & & 0.01    & & 0.22         \\ \hline
\end{tabular}
\end{center}
\end{table}

\subsubsection{\label{merger}Detecting SMBH in the Merger phase}

The merger phase can be related to the Fourier frequency interval
associated to the merger (lower) frequency $f_{m}$ all the way to the
ring-down frequency $f_{rd}$. During this evolutionary phase a
significant amount of GW energy, comparable to that emitted during the
ring-down phase, is radiated.

One of the main issues regarding the characterization of the merger
phase is the determination of the frequency $f_{m}$. We do not address
this issue here, and refer the reader to Ajith {\it
  et al.}  \citep{Ajith2008}, where it is shown that $f_{m} \simeq
0.04/M$ for equal mass BBH systems. Also in \citep{Ajith2008} it is
shown that there exist a reliable analytic approximation for the
merging-phase waveforms derived numerically. The resulting expression
for the radiated energy spectrum is equal to
\begin{equation}
\frac{d E}{d f} = \frac{1}{3} \pi^{2/3} \mu M^{2/3} \frac{f^{2/3}}{f_{m}}
\label{dEdfmerger}
\end{equation}
which is valid for two non-spinning and equal mass black-holes.

Finally, substituting the above equation into equation (\ref{snr}) we
get
\begin{equation}
\langle SNR^2 \rangle = \frac{[(1 + z)M]^{5/3}}{30 \pi^{4/3} f_{m} D_{L}(z)^2} \,
\int_{f_{min}/(1+z)}^{f_{m}/(1+z)}  \, \frac{df}{f^{4/3} S_h(f)}.
\label{snr2m}
\end{equation}

\begin{figure}
\begin{center}
\includegraphics[width=10cm]{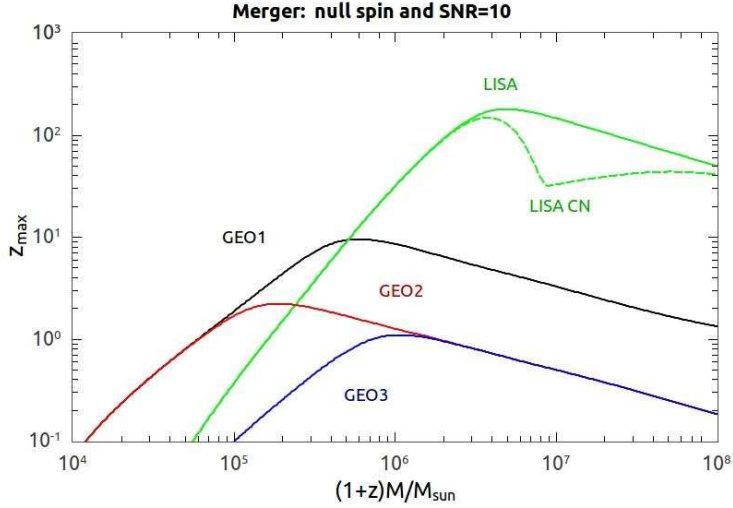}
\caption{
  $z_{max}$ as a function of the SMBH mass M for $SNR=10$, and null
  spin, for the different GEOGRAWI configurations.  Included for
  comparison are the results for LISA and LISA CN.}
\label{merg}
\end{center}
\end{figure}

Following the same prescription as for the ring-down case, we can then
calculate the redshift, $z_{max}$, below which a given SMBH of mass
$M$ can be detected with a given SNR. In Figure \ref{merg} we plot
$z_{max}$, for the three different GEOGRAWI configurations, as a
function of the mass $M$ of the SMBH system. As before we have
included the results valid for LISA and LISA CN. Also in this case the
GEO1 configuration represents an interesting alternative to LISA as
merging SMBHs could be seen at redshifts as high as $z = 10$.

For sake of curiosity, we compared our results against those
corresponding to a ``white'' spectrum \citep{flhu}. We found that the
two would agree quite well by assuming a white spectrum corresponding
to an energy emission efficiency of about $1\%$ (i.e.  $1\%$ of the
total mass $M$ of the BBHs is converted to gravitational radiation).

\subsubsection{\label{inspiral}Detecting SMBH in the inspiral phase}
\label{spiral}

In this section we finally analyze the inspiraling phase of a binary
system containing SMBH. In this case the signals frequency components
will fall into the region of the band where $ f < f_{m}$, which is
accessible to all the GEOGRAWI configurations considered here.

The energy spectrum for the inspiraling phase is given by the well
known formula \citep{Tho87}

\begin{equation}
\frac{d E}{d f} = \frac{1}{3} \pi^{2/3} \mu M^{2/3} f^{-1/3} \
\label{dEdfinspiral}
\end{equation}
and, for the particular case of binary systems whose components have
equal mass, the SNR becomes equal to

\begin{equation}
\langle SNR^2 \rangle = \frac{[(1 + z)M]^{5/3}}{30 \pi^{4/3} D_{L}(z)^2} \,
\int_{f_{min}/(1+z)}^{f_{m}/(1+z)}  \, \frac{df}{f^{7/3} S_h(f)}.
\label{snr2}
\end{equation}

\begin{figure}
\begin{center}
\includegraphics[width=10cm]{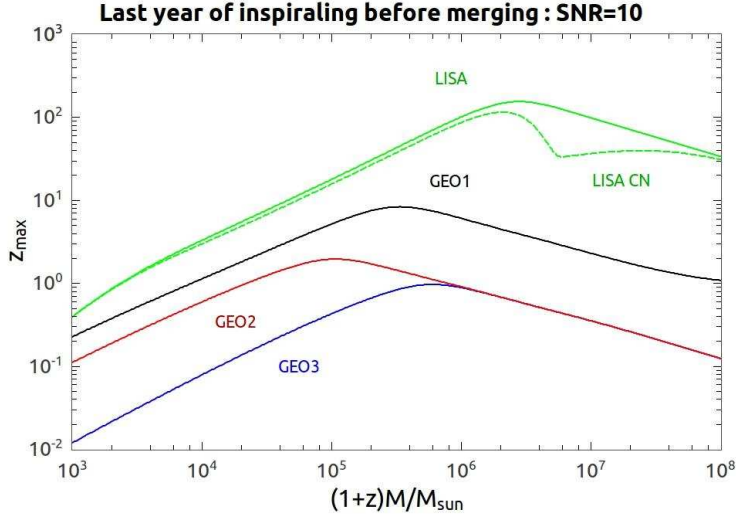}
\caption{
  $z_{max}$ as a function of the SMBH mass M for $SNR=10$ for the
  different GEOGRAWI configurations. Also shown for comparison are the
  curves corresponding to LISA and LISA CN.}
\label{insp}
\end{center}
\end{figure}

Note that now, in order to obtain $z_{max}$ as a function of $M$, we
had to treat the $SNR$ and $f_{min}$ as free parameters. As before, we
take the $SNR=10$. The value of $f_{min}$ is chosen in such a way that
the signal is integrated (observed) over the last year of inspiraling.
For binary systems whose components have equal masses it is easy to
show that $f_{min}$ is given by the following expression

\begin{equation}
f_{min}(T) = \left\{ f_{m}^{-8/3} + \frac{64}{5}
\pi^{8/3} [M(1+z)]^{5/3} T \right\}^{-3/8},
\end{equation}
where $T$ is the time before merging. In Figure \ref{insp} we plot
$z_{max}$ as a function of the mass of the system of SMBHs for ${\rm
  SNR}= 10$ and for one year of observation.

Although Fig. (\ref{insp}) shows that, at an SNR=10, the maximum
redshift achievable by LISA is consistently larger than that for GEO1,
also indicates that the maximum redshift at which the coalescence
phase can be observed by GEO1 is equal to about 5.

In summary, LISA can reach a larger redshift as compared to any
  GEOGRAWI configuration independently of the SMBH masses.

\section{Conclusions}
\label{Conclusions}

The main goal of this article was to analyze the performance and some
scientific capabilities of a geostationary gravitational wave
interferometer. We have done so by estimating its sensitivities to
gravitational radiation when operated under three different onboard
subsystem configurations, and analyzed the kind of gravitational wave
sources it will be able to detect and study in the sub-Hertz band. We
found our proposed Earth-orbiting detector less sensitive than the
Laser Interferometer Space Antenna mission by about a factor of
seventy in the lower part of its accessible frequency band ($10^{-4} -
2 \times 10^{-2}$ Hz). On the other hand, in the higher part of the
band ($2\times 10^{-2} - 10$ Hz), it was shown that GEOGRAWI has
observational capabilities that are complementary to those of the LISA
mission. In relation to this point, we emphasized that our assumed
noise performance in the high-frequency region of the accessible band
requires additional instrumental capabilities over those baselined for
the LISA mission. Although current experimental evidence
\cite{Jordan2006,Dan1,Dan2} indicates that they will be available by
the time such a mission will be considered, a more in-depth analysis
of these challenges is beyond the scope of this paper.

In the case of binary systems containing SMBHs, our analysis has shown
that the GEO1 interferometer could observe a great number of events
per year. This is direct consequence of its good sensitivity in the
frequency region ($2\times 10^{-2} - 10$ Hz), where lighter (and
larger in number) black-holes radiate in the merger and ring-down
phases. Consistently we have also shown that these same systems
  will be detectable by LISA with a SNR larger than that achievable by
  GEO1 during the inspiraling phase (see Fig. \ref{insp}).

Since most of the SNR from the coalescence phase of a binary system
containing SMBHs is achieved after about a week of integration time
before coalescence, because GEOGRAWI will make most of its
observations in the high-part of its accessible frequency band, and
because during a time period of one week the amplitude modulation of
the GEOGRAWI response will result into a frequency modulation of the
observed GW signal that depends on the source location in the sky, one
can argue that GEOGRAWI should achieve very good accuracies in the
reconstructed GW source parameters of SMBH with masses that are in the
range $10^4 - 10^6 \ M_\odot$. It is worth noting that McWilliams
\cite{PCOS,McWillians} has performed such an analysis for the GADFLI
mission concept (which is similar to GEOGRAWI). He showed that the
precision in the estimation of the parameters achievable by GADFLI
(which is analogous to our GEO1 configuration) is somewhat better than
that for LISA. Although these results imply that GEO1 should achieve
comparable accuracies in estimating the parameters characterizing
SMBHB, it is our intention to quantitatively study this problem in a
forthcoming article.

Since the high-part of the accessible frequency where GEOGRAWI
achieves its best sensitivity is at around $1$ Hz, we will investigate
also its ability to perform low-latency searches of burst events.
Mainly due to its capability of continuously transmitting real-time
data to the ground (by being geostationary), it should be able to
trigger simultaneous searches for electromagnetic counter-parts.

\section*{Acknowledgments}

M.T. acknowledges financial support from FAPESP through its Visiting
Professorship program at the Instituto Nacional de Pesquisas Espaciais
(I.N.P.E.) in Brazil (grant \# 2011/11719-0). JCNA and ODA would like
to thank CNPq and FAPESP for financial support. MESA would like to thank FAPEMIG for financial support (grant APQ-00140-12). Last, but not least,
we would like to thank the referee for his (her) useful suggestions
and criticisms.

\appendix

\section{Noise analysis}
\label{noise}

In this Appendix we provide an analysis of the noise sources affecting
the heterodyne measurements performed by the geostationary GW
interferometer GEO1. Since some of its onboard instrumentation is
similar to that of the LISA mission, it imposes the most stringent
noise-performance requirements on the subsystems affecting the one-way
heterodyne measurements.  Furthermore, as it will be discussed below,
the noise performance in the high-frequency region of the accessible
band requires additional instrumental capabilities over those
baselined for the LISA mission.  Although current experimental
evidence
\cite{Esteban2011,Pointing1997,Vinzenz,Daniel2006,Jordan2006,Dan1,Dan2}
induces us to believe that they will be available by the time such a
mission will be further analyzed, it is clear that these potential
challenges are well beyond the scope of this paper.

Our analysis will also rely on two LISA study documents as guiding
references \citep{PPA98,LISA1}.  There it was shown that there exist
two main categories of sensitivity-limiting noise sources:
\begin{enumerate}[(I)]
\item {Acceleration noises}
\item {Optical-Path noises}
\end{enumerate}
\noindent
The {\it acceleration noises} are due to residual forces acting on the
proof-masses of the Gravitational Reduction System (GRS), and result
into Doppler fluctuations into the heterodyne measurements. Their
magnitudes are most prominent in the low-part of the accessible
frequency band ($10^{-4} - 10^{-2}$ Hz) and depend on the (i) adopted
GRS design, (ii) the spacecraft design, and (iii) the mission
trajectory and space environment within which the spacecraft will be
operating. For these reasons the classification of the acceleration
noises is a multi-parameters problem and a very challenging one.  In
the case of the LISA mission this has already been studied extensively
and it will be finalized through the LISA Pathfinder experiment
\citep{Vitale1,Vitale2}. This is a ESA mission aimed at testing the
noise modeling, classification, and performance of the LISA GRS
system.

The LISA GRS was envisioned to have two cubic proof-masses onboard
each spacecraft and whose positions relative to the spacecraft were
measured with electrostatic (capacitative) readouts that were part of
their caging systems. In recent years, however, it has been argued
that a single, spherical proof-mass design (whose position relative to
its enclosing cage is measured optically) could be used instead,
providing significant simplifications of the GRS and of the optical
bench where the heterodyne measurements are performed \citep{Dan2005}.

Although a single, spherical proof-mass GRS implementation might
result into a better noise performance than that with two cubic masses
\citep{Dan2005}, in our analysis we will assume GEO1
to rely on a spherical GRS design whose performance is similar to that of the LISA
GRS, i.e. with a square-root of the acceleration spectral density
equal to a value of $3 \ \times 10^{-15} \ {\rm m/s} \ {\rm
  Hz}^{-1/2}$ over the accessible frequency band. In ultimate analysis
it will be the LISA Pathfinder experiment that will show us whether
the types and magnitudes of the noise sources affecting the GRS
performance are what we expect them to be.

The primary noise sources within the second category, i.e. {\it
  Optical-path noises}, are much easier to identify and are most
prominent in the higher part of the accessible frequency band. They
can be summarized as \citep{PPA98,LISA1}

\begin{itemize}
\item {Shot-noise at the photo-detectors,}
\item {Residual laser phase noise in the TDI observables,}
\item {Laser beam-pointing fluctuations,}
\item {Phase-meter noise,}
\item {Master clock noise,}
\item {Scattered light effects.}
\end{itemize}
\noindent
The {\it Shot-Noise} is a fundamental noise limitation to the
sensitivity of a laser interferometer GW detector in the high-part of its
accessible frequency band. It affects the one-way Doppler measurements
right at the photo-detector where two laser beams are made to
interfere, and it leads to the following spectral density of relative
frequency (Doppler) fluctuations \citep{PPA98}
\begin{equation}
S_{\rm shot} (f) = \frac{h f^2}{\nu_0 \ P_{\rm avail}} \ .
\label{shot}
\end{equation}
In Eq. (\ref{shot}) $h$ is the Planck constant, $\nu_0 \equiv 3.0
\times 10^{14} {\rm Hz}$ is the nominal laser frequency, $f$ is the
Fourier frequency, and $P_{\rm avail}$ is the effective power
available at the receiving photodetector. By assuming the same optics,
optical telescope size, laser power, and photodetector quantum
efficiency as those of LISA, from Eq. (\ref{shot}) we conclude that
the amplitude of the relative frequency fluctuations due to shot noise
affecting the GEO1 interferometer
scale down linearly with the interferometer armlength.

The {\it Residual laser phase noise} represents the ``left-over'' of
the laser noise in the TDI observables after the TDI algorithm is
applied to the one-way Doppler measurements by properly time-shifting
and linearly combining them. It is primarily due to the finiteness in
the accuracy by which the armlengths are known, and is proportional to
the amplitude of the laser frequency fluctuations. The relationship
between the magnitude of the spectrum of the residual laser frequency
fluctuations, $S_{\Delta C} (f)$, and the armlength accuracies,
$\delta L_i \ , i = 1, 2$, is given in \citep{TA98} for the unequal-arm
Michelson interferometer TDI combination and has the following form
\begin{eqnarray}
S_{\Delta C} (f) & = & 64 \pi^2 f^2 S_C(f) \left\{
\delta L_1^2 \sin^2(2 \pi f L_2) + \delta L_2^2 \sin^2(2 \pi f L_1)
\right.
\nonumber
\\
& - & \left. \delta L_1 \delta L_2 [\sin^2(2 \pi f L_1) + \sin^2(2 \pi f L_2) -
\sin^2(2 \pi f (L_2 - L_1))]\right\} \ .
\label{laserTrue}
\end{eqnarray}
In the long-wavelength limit it is easy to show from the above
equation that the amplitude of the residual laser frequency
fluctuations scale linearly with the armlength of the interferometer
(see Eq. (3.18) of \citep{TA98}). On the other hand, this scaling no
longer exists at higher-frequencies and, in order to maintain this
noise source negligible in the GEO1 noise budget, a higher level of
accuracy in the knowledge of the armlengths is required. Since the
GEO1 configuration will have a shot noise amplitude smaller than that
of LISA by about a factor of $70$, by measuring the GEO1 armlength
with an accuracy that is seventy times better than that of LISA we
will make this noise source negligible in the GEO1 TDI combination
$X$. Given that the required LISA armlength accuracy for $X$ has been
estimated to be equal to about $30$ meters \citep{TA98}, we conclude
that the armlength accuracy needed for GEO1 should be of about $40$
centimeters. Such a level of laser ranging accuracy has already been
demonstrated experimentally \citep{Esteban2011} at a much lower
receiving laser power. This means that, in the case of the GEO1
mission, the achievable armlength accuracy will be smaller than $40$
cm, further reducing the GEO1 residual laser frequency noise.

The {\it Laser beam-pointing fluctuations} are due to distortions in
the transmitted laser beam wavefront that appear at the receiving
spacecraft as additional frequency fluctuations generated by small
pointing fluctuations from the transmitting spacecraft. In the case of
the LISA mission, whose pointing fluctuations specifications were
required not to be greater than $6 \ {\rm nrad} \ {\rm Hz}^{-1/2}$, it
was shown \citep{Pointing1997} that a way for substantially reducing
pointing-induced frequency fluctuations was to rely on the light
received from the far spacecraft to sense the orientation of the
receiving spacecraft. An ingenious way for doing so \citep{Pointing1997} is by
sampling some of the incoming light on a CCD array. By detecting the
position of the beam on the CCD it is then possible to deduce the
alignment of the spacecraft and implement it by proper control
signals. This measurement is shot-noise limited and, in the case of
the LISA mission, it was proposed to implement it by extracting
$10$ percent of the power of the incoming laser light. This
resulted into a pointing noise of about $0.5 \ {\rm nrad} \ {\rm
  Hz}^{-1/2}$, more than a factor of $10$ better than that
specified for LISA. Since the shot-noise level associated with this
pointing measurement technique scales down linearly with the armlength
(by being inversely proportional to the square-root of the received
laser power) we conclude that the pointing shot-noise limit for
GEO1 will scale down linearly with the armlength from the level
estimated for LISA.

The {\it Phase-meter} is the subsystem that measures the difference
between the phase of the incoming laser beam and that from the local
laser, and in the process it introduces an additional phase noise in
the one-way Doppler measurements.  Although the noise it generates
does not scale with the armlength, it does depend on the magnitude of
the relative frequency offset between the transmitted and received
frequencies of the two interfering laser beams \citep{Vinzenz}.  As an
example, it has been shown that the frequency stability of the
phase-meter to be flown on the LISA Pathfinder mission (which has an
heterodyne frequency range of about a few kHz) will be several orders
of magnitude better than that of the phase-meter for the LISA mission
\citep{Vinzenz,Daniel2006}. It is for this reason that, in order to
compensate for the Sun and the Moon gravitational perturbations on
each spacecraft and maintain orbital stability and small
inter-spacecraft relative velocities, each spacecraft will implement
``station-keeping'' \citep{Soop,Drift}.  Since this operation will
need to be performed about once per week \citep{Kamel} in order to
maintain the spacecraft relative velocities smaller than about a few
decimeters per second, it will not disrupt significantly the science
data acquisition. As an additional note, GEO1 will also rely on
molecular iodine laser frequency stabilization, which has been shown
to provide laser frequency stability superior to that achievable by
optical cavity stabilization and a laser frequency accuracy at the
level of $1$ to $2$ kHz \citep{Jordan2006}.

The {\it Master clock noise} affects the LISA measurements because it
is used for removing the large Doppler beat-notes (as large as $20$
MHz) affecting the Doppler measurements.  Since GEO1 will not be
subject to large orbital perturbations as those affecting LISA and it
will rely on station-keeping in order to maintain a ``stationary''
configuration, the noise due to the onboard master clock for
performing the heterodyne measurements will be negligible as it is
proportional to the magnitude of the beat note at the photodetector
\citep{TEA2002}. This will result into a significant simplification of
the designs of the phase-meter and transmitting modules as modulations
of the transmitting and receiving beam (for implementing the
clock-noise cancellation scheme) will not be needed.

{\it Scattered Light effects} are due to spurious light signals
generated from the interaction of the received main laser beam with
the optical telescope, beam-splitter, and other optical components of
the optical bench the beam interacts with before being made to
interfere with the light of the local laser. Although this noise
source for quite sometimes has been thought to be unavoidable, in
recent years new developments in laser metrology have indicated that
scattered light noise can be suppressed to sub-picometer levels in the
frequency band of interest to space-based GW interferometers. In one
such laser metrology technique, called ``Digital Interferometry'' (DI)
\citep{Dan1,Dan2}, laser light is phase modulated with a pseudo-random
noise (PRN) code that time stamps the light, allowing isolation of
individual propagation paths based on their times of flight through
the system. After detection, the PRN phase shift is removed by
decoding with an appropriately delayed copy of the code. By matching
the decoding delay to the propagation delay the signal is coherently
recovered, allowing phase measurements to extract displacement
information at resolutions much smaller than the laser wavelength.
This matched-delay filtering of DI allows isolation of reflections
from an object at a particular distance, such as the optical
components giving origin to the scattered light noise.

In summary, the magnitude of the relative frequency fluctuations due
to the shot-noise and the laser beam-pointing fluctuations scale
linearly with the armlength, as it follows from the considerations
leading to equation (4.4) at page 82 of \citep{PPA98}, and sections 3
at page 1576 of \citep{Pointing1997} respectively. Although the
frequency fluctuations due to the residual laser phase noise, the
scattered-light noise, and the phase-meter noise do not follow a
similar scaling, we have shown that they can be made negligible within
the GEO1 noise budget. This is because of a demonstrated better
armlength accuracy \citep{Esteban2011}, of digital interferometry as
it suppresses scattered light effects, and adoption of a simpler
phase-meter design \citep{Vinzenz} that takes advantage of much
smaller ``beat-note'' frequencies among the lasers. The latter is made
possible through a combination of laser iodine stabilization
\citep{Jordan2006} together with spacecraft station-keeping maneuvers
\citep{Soop}.

\section{Coalescence rate}
\label{appendix}

In this Appendix we discuss the differential coalescence rate per
logarithm frequency interval as this can be related to the
differential event rate associated, for example, with the ring-down
phase shown in Fig. \ref{Differ}.  For a more detailed analysis
regarding the coalescence rate and the corresponding cosmological
simulations, we refer the reader to Refs. \citep{fill} and
\citep{fill0} respectively.

The differential coalescence rate is obtained from an appropriate
combination of Eqs. (\ref{erobs}) and (\ref{frd}) and an integration
in a given redshift interval, namely $0 < z < z_{initial}$, where
$z_{initial}$ refers to a value selected at the beginning of the
simulation.

\begin{figure}
\begin{center}
\includegraphics[width=8cm, angle=-90]{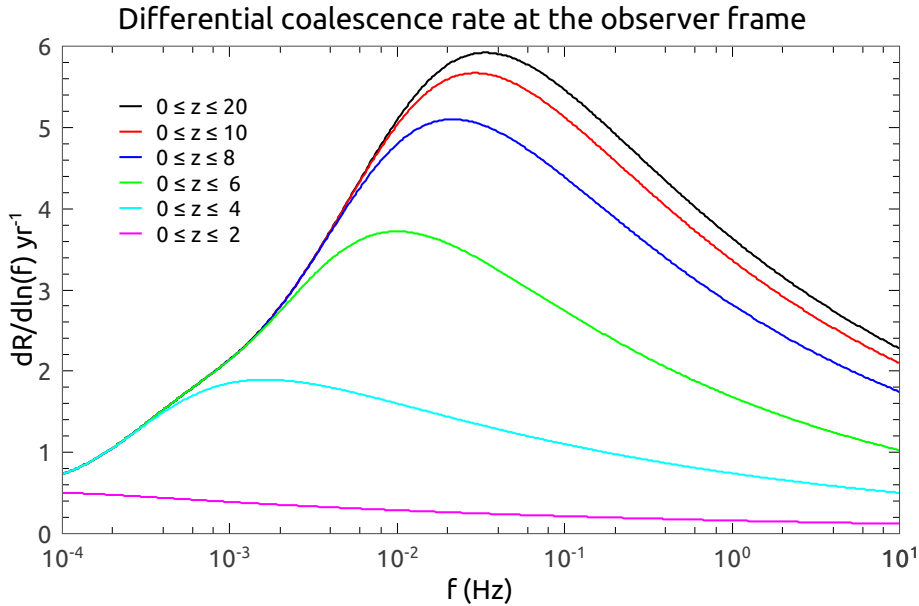}
\caption{Differential coalescence rate as a function of the Fourier
  frequency and for different interval of integration in redshift.}
\label{dcoalrate}
\end{center}
\end{figure}

Fig. \ref{dcoalrate} shows the differential coalescence rate per
logarithmic frequency interval as a function of the frequency and for
different intervals of integration in the redshift. We have not
included curves corresponding to values of the redshift higher than
$20$ because they all ``saturate'' to the curve corresponding to this
value of $z_{initial}$.

Another interesting information concerning the study by Filloux {\it
  et al.}  \citep{fill} refers to the coalescence rate at the observer
frame, which can be obtained by integrating the differential
coalescence rate. After performing such an integration we find a
coalescence rate of $\sim $ 43 yr$^{-1}$ (associated with the
  ring-down phase), which should be regarded as the maximum event
rate a given gravitational wave antenna might detect.  Our study has
actually shown that LISA and GEO1 should be able to detect a large
fraction of this rate.

\begin{figure}
\begin{center}
\includegraphics[width=8cm]{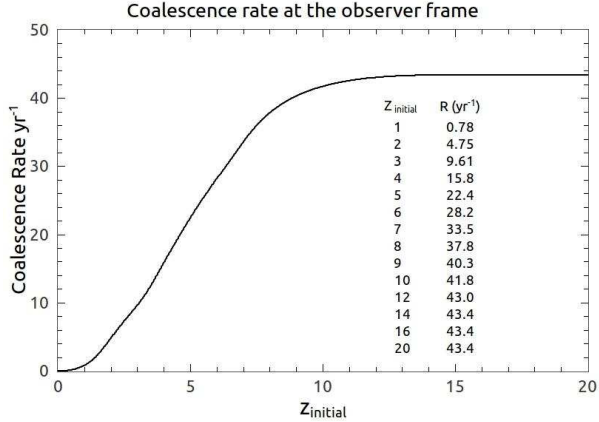}
\caption{Coalescence rate as a function of $z_{initial}$. Note how
  this function starts to plateau at $z_{initial} \approx 12$.}
\label{coalrate}
\end{center}
\end{figure}

Implicit in the integration just mentioned, there is an integration in
the redshift interval $0 < z < z_{initial}$, where $z_{initial}$ was
taken to be equal to $60$ in the present study.  In practice, however,
there is a saturation around $z \sim 12$, as can be seen in Fig.
\ref{coalrate}, where the cumulative event rate as a function of
$z_{initial}$ is plotted and tabulated.  Note that half of the
coalescence rate comes from events occurring at $z > 5$.

\end{document}